\documentclass[aps,prd,10pt,nofootinbib,twocolumn,eqsecnum,showpacs,showkeys,superscriptaddress,preprintnumbers,altaffilletter,floatfix]{revtex4-2}
\DeclareUnicodeCharacter{0327}{\c{}}
\DeclareUnicodeCharacter{202F}{ }
\usepackage[utf8]{inputenc}
\usepackage[T1]{fontenc}
\usepackage{graphicx}
\usepackage{amsmath}
\usepackage{multirow}
\usepackage{hyperref}
\usepackage{soul}
\usepackage{graphicx}
\usepackage{dcolumn}
\usepackage{amssymb}
\usepackage{amsfonts}
\usepackage{physics}
\usepackage{amsbsy}
\usepackage{color}
\usepackage{rotating}
\usepackage[english]{babel}
\usepackage{multirow}
\usepackage{float}

\newcommand{\be}{\begin{equation}}
\newcommand{\ee}{\end{equation}}
\newcommand{\bea}{\begin{eqnarray}}
\newcommand{\eea}{\end{eqnarray}}
\hypersetup{
    colorlinks=true,
    linkcolor=blue,
    filecolor=magenta,
    urlcolor=cyan,
    citecolor=cyan
}

\begin{document}

\title{Mass-to-Horizon Entropic Cosmology: A Unified Thermodynamic Pathway to Cosmic Acceleration}
% \title{Mass-to-Horizon Entropic Acceleration: A Fresh Perspective on the Origin of Dark Energy}
\author{Tomasz Denkiewicz}
\email{tomasz.denkiewicz@usz.edu.pl}
\affiliation{Institute of Physics, University of Szczecin, Wielkopolska 15, 70-451 Szczecin, Poland}
\author{Hussain Gohar}
\email{hussain.gohar@usz.edu.pl}
\affiliation{Institute of Physics, University of Szczecin, Wielkopolska 15, 70-451 Szczecin, Poland}

\date{\today}
\begin{abstract}
We investigate the observational tests of generalized mass-to-horizon entropic cosmology by incorporating large-scale structure growth data in addition to purely geometric probes. The theoretical framework is constructed from a generalized mass-to-horizon scaling relation, $M \propto L^n$, which implies a corresponding generalized entropic functional $S_n \propto L^{n+1}$. Within this setting, cosmic acceleration arises as an emergent phenomenon driven by an entropic force acting on the cosmological horizon. While earlier studies demonstrated that these entropic cosmologies can reproduce the background expansion history of the standard $\Lambda$CDM model, here we present a comprehensive observational analysis that jointly employs Pantheon+ Type Ia supernova data with SH0ES calibration, DESI DR2 baryon acoustic oscillation measurements, cosmic microwave background (CMB) distance priors, and a suite of cosmological structure growth observations. A Bayesian model comparison indicates that the entropic models are statistically preferred over the conventional $\Lambda$CDM scenario, thereby providing strong support for an entropic origin of the observed late-time cosmic acceleration in place of a fundamental cosmological constant.
\end{abstract}
\maketitle
\section{Introduction}
Recent evidence indicates that the universe is experiencing an unprecedented rate of expansion \cite{Perlmutter:1998np,Riess:1998cb,Riess:2004nr,Riess:2006fw,Pan-STARRS1:2017jku,Hinshaw:2012aka,Aghanim:2018eyx,Alam:2020sor}. A multitude of studies have been conducted to explore this phenomenon \cite{Li:2011sd,Bamba:2012cp}. Dark energy, characterized by its unique attributes that influence the Einstein field equations, has been suggested as a possible explanation for this acceleration. Nonetheless, models of dark energy encounter certain discrepancies when confronted with observational data and present theoretical challenges \cite{Bull:2015stt}. Consequently, numerous alternative approaches have been explored to comprehend the underlying physical mechanisms driving the accelerated expansion.
Among the promising frameworks under consideration is entropic acceleration, or entropic cosmology\footnote{It is important to distinguish entropic cosmology from entropic gravity \cite{Verlinde:2010hp}. In entropic cosmology, general relativity is assumed to hold, and additional entropic force terms are introduced into the field equations, whereas entropic gravity interprets gravity itself as an emergent force.}, as proposed in \cite{Easson:2010av,Easson:2010xf}. In this approach, terms associated with entropic forces—motivated by boundary contributions to the Einstein–Hilbert action (see, for instance, the detailed derivations in \cite{Garcia-Bellido:2021idr,Arjona:2021uxs,Garcia-Bellido:2024qau,Garcia-Bellido:2025hji,Calderon:2025dhj} in the context of general relativistic entropic acceleration) and further supported by the holographic principle \cite{Susskind:1994vu,tHooft:1993dmi}—are incorporated into the Einstein field equations. These entropic force contributions are postulated to drive the present cosmic acceleration and to provide a unified entropic explanation for both early- and late-time accelerated expansion. From a holographic standpoint, it is conjectured that the information encoded on the boundary of the universe induces an entropic force and an associated effective negative pressure, which may account for the observed rapid expansion of the universe.

Assuming the validity of the holographic principle, the earliest models of entropic cosmology employ the Hawking temperature \cite{Hawking:1974rv} and Bekenstein entropy \cite{Bekenstein:1973ur} defined on the boundary in order to derive the entropic force terms. However, these original formulations face serious difficulties in reproducing the transition between accelerated and decelerated expansion phases and fail to provide an adequate fit to observational growth-of-structure data \cite{Basilakos:2012ra,Basilakos:2014tha}. These shortcomings have been mitigated by phenomenologically introducing additional terms beyond those originally proposed in \cite{Komatsu:2012zh,Komatsu:2013qia,Komatsu:2014lsa,Komatsu:2014vna,Komatsu:2014ywa,Komatsu:2015nga,Komatsu:2015nkb, Gohar:2020bod}. Recently, new generalized entropic-force cosmological models were developed in \cite{Gohar:2023lta,Gohar:2023hnb} by one of the authors of the present work, which successfully resolve the main limitations of the original scenarios. It has been demonstrated that these generalized models are statistically indistinguishable from the standard Lambda Cold Dark Matter ($\Lambda$CDM) cosmological model in terms of Bayesian evidence, and they yield parameter estimates that are effectively identical to those of the standard scenario. This provides the first concrete realization of a fully entropic framework underlying dark energy. However, in \cite{Gohar:2023lta}, the growth of cosmic structures was not analyzed, as the primary objective was to establish an entropic origin for dark energy. In this work, we examine the growth of cosmic structures within the framework of generalized entropic cosmologies, with the goal of addressing the evolution of structure formation and incorporating the latest observational datasets to constrain the model parameters

One of the foundational components of entropic cosmology is the specification of the entropic force contributions, which are intrinsically determined by the particular entropy functional associated with the chosen cosmological horizon. A broad spectrum of entropy proposals have been proposed, inspired by developments in nonextensive statistical mechanics \cite{Tsallis:1987eu, tsallisbook}, thermodynamics \cite{Tsallis:2012js,Zamora:2022cqz, Zamora:2022sya}, quantum gravity \cite{Rovelli:1996dv, Meissner:2004ju}, and various phenomenological frameworks \cite{Barrow:2020tzx}. Recent investigations have demonstrated that many of these generalized entropy definitions, which extend beyond the standard Bekenstein entropy, become thermodynamically inconsistent when combined with the Hawking temperature \cite{Gohar:2023lta, Cimdiker:2022ics, Nojiri:2021czz, Gohar:2023hnb}, thereby generating tensions within holographic scenarios. These inconsistencies have been systematically examined in the context of entropic cosmology in \cite{Gohar:2023hnb, Gohar:2023lta}, and in the context of black hole thermodynamics in \cite{Cimdiker:2022ics, Nojiri:2021czz}. 

In the holographic scenarios, thermodynamic consistency requires that any entropy $S$ and 
temperature $T$ attributed to a holographic horizon, when inserted into the Clausius relation $dE = T\, dS$, must preserve the identification of the energy $E$ with 
the mass $M$. In most generalizations that go beyond the Bekenstein entropy, the strategy is to deform or extend the Bekenstein entropy while keeping the standard Hawking 
temperature unmodified. The central question is whether these generalized entropy functionals, when used in conjunction with Hawking temperature, still satisfy 
the conditions of thermodynamic consistency. A closely related issue is whether the adoption of the Hawking temperature remains theoretically justified once such non-Bekenstein entropy forms are introduced. Moreover, it has been demonstrated that generalized entropy functionals fail to satisfy the consistency condition when combined with the Hawking temperature \cite{Nojiri:2021czz,Cimdiker:2022ics}. Although one may attempt to modify the Hawking temperature, as it is done in \cite{Cimidiker:2023kle}, to restore consistency, such modifications lack a clear justification from quantum field theory. 

A key but often overlooked element in holographic cosmology is the linear mass--horizon relation (MHR), $M = \gamma c^{2} L / G$, where $\gamma>0$ is a dimensionless parameter. Although commonly used---typically implicitly---when invoking the Clausius relation and the first law of thermodynamics, its role has rarely been stated explicitly. This assumption was first clarified in our earlier work~\cite{Gohar:2023hnb, Gohar:2023lta} and later generalized in~\cite{Gohar:2025yfx}.
For a Schwarzschild black hole, the MHR reduces to $M = c^{2} r_{+} / (2G)$ with $r_{+}$ the horizon radius, a result that follows directly from black-hole geometry. In cosmological holographic applications, however, this linear relation is an assumption, yet an essential one for thermodynamic consistency. In particular, combining the Bekenstein entropy with the Hawking temperature requires this relation to preserve the holographic consistency of the thermodynamic framework.
Despite its significance, the necessity of the MHR is seldom mentioned in the literature, even though standard cosmological applications implicitly rely on it.

Motivated by the aforementioned concerns and questions, and in order to restore thermodynamic consistency between generalized entropy formalisms and the Hawking temperature, one of the authors of this work introduced in \cite{Gohar:2023lta} a generalized mass–horizon relation\footnote{For a more general relation, see \cite{Gohar:2025yfx}.}.
\be
M = \gamma \frac{c^2}{G} L^n,
\ee
where $c$ denotes the speed of light, $G$ is Newton’s gravitational constant, and $\gamma$ and $n$ are nonnegative free parameters. This relation is then employed to derive the generalized mass–to-horizon entropy $S_n$
\be
S_n = \frac{2\pi k_B n \gamma}{(n+1) l_p^2} L^{n+1}, \label{S_m}
\ee
where $\gamma$ has dimensions of $[\text{length}]^{1-n}$, $l_p$ is the Planck length, and $k_B$ is the Boltzmann constant. Notably, $S_n$ reproduces the standard Bekenstein entropy, as well as the Tsallis–Cirto entropy, Barrow entropy, and Tsallis–Zamora entropy for appropriate choices of the parameters $n$ and $\gamma$ (see Refs.~\cite{Gohar:2023lta, Gohar:2025yfx} for these parameterizations). More importantly, with this mass-to-horizon relation, the new definition of $S_n$ is thermodynamically consistent with the Hawking temperature from a holographic perspective. Consequently, all entropy functionals that can be obtained as particular limits or specializations of $S_n$ are thermodynamically consistent, provided that the parameter $n$ in the mass-to-horizon relation is chosen in accordance with the specific entropy definition under consideration.

Within the entropic cosmology framework, these questions have been examined in \cite{Gohar:2023hnb}, where we demonstrated that, upon an appropriate modification of the Hawking temperature to restore the holographic consistency, all entropic cosmological models based on different entropy functionals become equivalent to the conventional entropic cosmology derived from Bekenstein entropy and the Hawking temperature. This result implies that, regardless of the specific entropy functional employed, one cannot observationally distinguish between the corresponding entropic force models, even though the various entropy definitions possess well-motivated theoretical justifications.
Subsequently, in \cite{Gohar:2023lta}, this observational degeneracy was revisited by introducing a generalized mass--to--horizon relation, which enabled a broader and thermodynamically consistent formulation of entropic cosmology. This generalized setup leads to a continuous family of cosmological scenarios characterized by an index $n$. Remarkably, the case $n=3$ reproduces exactly the behavior of a cosmological constant, thereby offering an entropic interpretation for its origin within this framework.

% Building upon this formulation, one can associate to each entropy functional $S_n$ a corresponding entropic force $F_n$ and pressure $p_n$ given by \cite{Gohar:2023lta}
% \begin{equation}
% F_n = -\gamma n \frac{c^4}{G}L^{\,n-1}, 
% \qquad
% p_n = -\gamma n \frac{c^4}{4\pi G}L^{\,n-3}.
% \label{eq:entropic_F_p}
% \end{equation}
% These quantities can be expressed in a fluid-like form by introducing an effective entropic energy density $\rho_n$, such that 
% %\begin{equation}
% $p_n = -c^2 n\,\rho_n$ ,
% %\label{eq:p_rho}
% %\end{equation}
% where 
% \begin{equation}
% \rho_n = \gamma\, \frac{c^{\,n-1}}{4\pi G}\,H^{\,3-n}.
% \label{rho_n}
% \end{equation}
% For $n=3$, the parameter $\gamma$ plays the role of a cosmological constant, recovering the standard dark energy behavior. More generally, for arbitrary $n$, the quantity $\rho_n$ represents an effective entropic fluid whose contribution can drive the accelerated expansion of the universe. This perspective provides a unified interpretation of dark energy--like phenomena as emerging from generalized entropic considerations. 

Concerning the growth of structures, early entropic cosmologies faced a persistent difficulty: in their simplest realizations, the canonical entropic-force contribution, typically scaling as $H^2$, failed to provide an adequate description of the linear growth of matter perturbations and was unable to fit Redshift Space Distortion (RSD) measurements of $f\sigma_8(z)$ without reverting to an expansion history effectively indistinguishable from $\Lambda$CDM \cite{Basilakos:2012ra, Basilakos:2014tha}. Subsequent attempts to generalize the underlying entropy, for instance through Tsallis–Cirto, R'enyi, or Tsallis–Zamora nonextensive forms, did not resolve this tension: when implemented within a thermodynamically consistent framework, these entropies become functionally equivalent to the Bekenstein–Hawking case and therefore inherit the same shortcomings regarding structure formation \cite{Gohar:2023hnb}. Phenomenological extensions based on dissipative or matter-creation terms were also explored \cite{Komatsu:2014lsa, Komatsu:2014vna}, but the observed growth could only be accommodated for very small creation/dissipation rates, with larger rates strongly suppressing the linear density contrast $\delta(a)$. Other nonextensive constructions, such as Tsallis/Barrow-type scenarios, were shown to qualitatively reproduce the observed $f\sigma_8(z)$ locus by modifying both the background and the effective clustering sector; however, most of these analyses relied on curve-overlays rather than fully fledged covariance-level MCMC fits to standard RSD compilations \cite{Sheykhi2022, Basilakos2024}. Likewise, Tsallis holographic dark energy (THDE) models, although explicitly confronted with $f\sigma_8$ data, generally remained statistically disfavored with respect to $\Lambda$CDM according to information-criterion-based model selection \cite{daSilva:2021TsallisPert}. In this context, the generalized Mass-to-Horizon Entropic Cosmology (MHEC) framework \cite{Gohar:2023lta} restores thermodynamic consistency and yields a background expansion compatible with $\Lambda$CDM, but a systematic and rigorous confrontation with cosmological growth data has so far been lacking; establishing whether the same entropic mechanism can also reproduce the observed evolution of structure formation is therefore a key open question that we address in this work.

The remainder of the paper is organized as follows. We first introduce the generalized entropic Friedmann equations, which include the entropic fluid $\rho_n$. In Section III, we present the formalism of linear perturbations. In Section IV, we describe the observational data sets and the methodology employed to constrain the parameters of our model. In Section V and VI, we report and discuss the main results. Finally, in Section VII, we summarize the principal conclusions of this study. 
%{\bf  Appendix \ref{appx} ... ... ... ...}

\section{Generalized Entropic Cosmological Models}\label{sec:II}
We consider a spatially flat, homogeneous, and isotropic Friedmann–Lemaître–Robertson–Walker (FLRW) universe and identify the Hubble horizon, defined by \(L = c/H\), as the relevant infrared cutoff scale. Motivated by the holographic principle, we endow this horizon with a Hawking temperature \(T = \hbar c/(2\pi k_B L)\) and a generalized mass-to-horizon entropy \(S_{n}\). The corresponding generalized entropic force, defined as \(F_{n} = -T\,dS_n/dL\), gives rise to an effective entropic pressure \(p_{n} = F_n/A\), which in turn modifies the Friedmann and acceleration equations, thus providing a possible mechanism for explaining the observed late-time cosmic acceleration. 

Within this framework, each entropy functional \(S_n\) can be associated with an entropic force \(F_n\) and pressure \(p_n\), which take the form \cite{Gohar:2023lta}
\begin{equation}
F_n = -\gamma n \frac{c^4}{G}L^{\,n-1}, 
\qquad
p_n = -\gamma n \frac{c^4}{4\pi G}L^{\,n-3}.
\label{eq:entropic_F_p}
\end{equation}
 These quantities can be cast in a fluid-like representation by introducing an effective entropic energy density \(\rho_n\), defined such that 
%\begin{equation}
$p_n = -c^2 n\,\rho_n$ ,
%\label{eq:p_rho}
%\end{equation}
with
\begin{equation}
\rho_n = \gamma\, \frac{c^{\,n-1}}{4\pi G}\,H^{\,3-n}.
\label{rho_n}
\end{equation}
For \(n = 3\), the parameter \(\gamma\) effectively plays the role of a cosmological constant, thereby reproducing the standard dark-energy behavior. More generally, for arbitrary \(n\), the quantity \(\rho_n\) describes an effective entropic fluid whose contribution to the total energy budget can drive an accelerated expansion of the universe. This formulation thus offers a unified interpretation of dark-energy–like phenomena as emergent effects arising from generalized entropic considerations at the Hubble horizon.

To systematically incorporate the contributions of these entropic forces into the Friedmann, acceleration, and continuity equations, we adopt the formalism developed in Ref. \cite{Gohar:2023lta}, with additional details and motivation provided in \cite{Komatsu:2014lsa,Komatsu:2013qia}. Within this framework, the modified cosmological equations for a multi–fluid system can be written in the following form:
\begin{align} 
&H^2 =\frac{8\pi G}{3}\sum_{i} \rho_i +  \frac{4 \pi G}{3} \rho_{n}  \left(3n-1\right) \,, \label{eq:Fa11}\\
&\frac{\ddot a}{a}=-\frac{4\pi G}{3}\sum_{i}\left(\rho_i+\frac{3p_i}{c^2}\right) +  \frac{4 \pi G}{3} \rho_{n}  \left(3n-1\right) \,, \label{eq:Aa11}\\
&\sum_i \dot{\rho}_i + 3 H \sum \rho_i (1+w_i) = -\frac{3n-1}{2}\dot{\rho}_n \, , \label{eq:continuity_Lambda_MLn_2}
 \end{align} 
where the subscript $i$ runs for matter and radiation. 
In a similar manner, we can write the continuity equations for the second case as 
\begin{align}
&\dot{\rho}_{n} = -A_{m} H^{2-n} \rho_m - A_{r} H^{2-n} \rho_r\,\, , \label{eq:continuity_Lambda_MLn_2_entropy}\\
&a \rho'_i + 3 \rho_i (1+w_{eff,i}) = 0 \, , \label{eq:continuity_Lambda_MLn_mat_rad_2} \\
& w_{eff,i} = w_i - \frac{3n-1}{6}A_i H^{1-n} \, , \label{eq:weff_2}
\end{align}
where $A_{m} = 4\pi G C^{\gamma}_n$ and $A_{r} = 16\pi G/3\,C^{\gamma}_n$ with $C^{\gamma}_{n} = (3-n)\gamma c^{n-1}/(4 \pi G)$.
We will solve numerically the system of differential equations in order to perform the comparison with data. Note that in our framework, the parameter $\gamma$ plays a crucial role in ensuring the viability of the model: if $\gamma$ is sufficiently small, an exchange of energy occurs between the matter/radiation sector and the entropic fluid, with the latter effectively behaving as a very slowly varying cosmological “constant”. For $n = 3$, the entropy scales as $L^4$, the mass scales with volume ($M \propto L^3$), the entropic densities $\rho_n$ is constant with $\gamma$ as effective cosmological constant, and the above entropic models are fully equivalent to $\Lambda$CDM.

\section{Linear Perturbation}\label{sec:lin_pert}
Within the (subhorizon) Newtonian approximation and in comoving coordinates $\vec{x}$, we decompose the fluid variables as
$\rho(\vec{x},t)=\bar\rho(t)+\delta\rho(\vec{x},t)$ with the dimensionless density contrast
$\delta(\vec{x},t)\equiv\delta\rho(\vec{x},t)/\bar\rho(t)$,
$p(\vec{x},t)=\bar p(t)+\delta p(\vec{x},t)$ with $\delta p(\vec{x},t)=c_s^2\,\delta\rho(\vec{x},t)$,
and the velocity field
$\vec v(\vec{x},t)=H\vec r+\vec u(\vec{x},t)=aH\vec x+\vec u(\vec{x},t)$,
where $\vec r=a(t)\vec x$ and $H\equiv\dot a/a$. We denote by
$\theta_i\equiv\nabla\!\cdot\!\vec u_i$ the divergence of the peculiar velocity of species $i$,
and consider first–order perturbations of the Newtonian potential $\Phi(\vec{x},t)=\phi(\vec{x},t)$.
With these conventions the continuity, Euler and Poisson equations read
\begin{equation}
\frac{d\delta_i}{da} 
+ \left[ \frac{3}{a}(c_{s,i}^2 - w_i) + \frac{1}{aH}\frac{dH}{da} \right]\delta_i
+ \frac{1+w_i}{a^2 H}\,\theta_i = 0,
\label{eq:continuity}
\end{equation}
\begin{equation}
\frac{d\theta_i}{da}
+ \left( \frac{1}{a} + \frac{1}{H}\frac{dH}{da} \right)\theta_i
= -\frac{1}{a^2 H}\left( \frac{c_{s,i}^2}{1+w_i}\nabla^2\delta_i + \nabla^2\phi \right),
\label{eq:euler}
\end{equation}
\begin{equation}
\nabla^2 \phi = 4\pi G a^2 \sum_j (1+3w_j)\,\bar{\rho}_j \delta_j,
\label{eq:poisson}
\end{equation}
where $w_i\equiv \bar p_i/\bar\rho_i$ may be time–dependent. Combining Eqs.~\eqref{eq:continuity}–\eqref{eq:poisson} yields the second–order evolution equation
(for primes denoting $d/da$)
\begin{widetext}
\begin{equation}
\begin{split}
\delta_i'' &+ \left[ \frac{3}{a}(c_{s,i}^2 - w_i) + \frac{3}{a} + \frac{1}{H}\frac{dH}{da} 
- \frac{1}{1+w_i}\frac{dw_i}{da} \right]\delta_i'  \\
&+ \Bigg[ \frac{3}{H}\frac{dH}{da}(c_{s,i}^2 - w_i) 
+ 3\left( \frac{dc_{s,i}^2}{da} - \frac{dw_i}{da} \right) 
+ \frac{6}{a}(c_{s,i}^2 - w_i) 
- \frac{3(c_{s,i}^2 - w_i)}{1+w_i}\frac{dw_i}{da} \Bigg]\delta_i
- \frac{c_{s,i}^2}{a^4 H^2}\nabla^2\delta_i \\
&= \frac{4\pi G (1+w_i)}{a^2 H^2} 
\sum_j (1+3w_j)\,\bar{\rho}_j \delta_j \, .
\end{split}
\label{eq:delta_second_order_a}
\end{equation}
\end{widetext}
We adopt an adiabatic barotropic closure for the standard fluids whose density contrasts we explicitly evolve,
namely cold matter and radiation, and in this (special) case we take $c_{s,i}^2=w_i(a)$ with $w_i\equiv \bar p_i/\bar\rho_i$.
For cold matter this implies $c_{s,M}^2\simeq 0$ ($w_M=0$), while for radiation $c_{s,R}^2=1/3$ ($w_R=1/3$),
as in the standard subhorizon Newtonian treatment of $\Lambda$CDM growth.
In our MHEC setup the interaction with the entropic sector affects the \emph{background} evolution through the source terms
$Q_i$ (equivalently encoded in the effective background equations of state $w_{M,\mathrm{eff}}(a)$ and $w_{R,\mathrm{eff}}(a)$),
but we do not interpret $w_{i,\mathrm{eff}}(a)$ as a pressure closure for perturbations.
Moreover, we do not evolve perturbations of the entropic contribution; on the subhorizon scales relevant for current growth data
we assume it remains smooth ($\delta_n\simeq 0$), so no separate sound-speed parameter is introduced for the entropic sector.
Within this effective-fluid approximation we neglect additional intrinsic (non-adiabatic) pressure perturbations of the
standard components, following the usual approach employed in entropic-force/nonextensive entropic cosmologies and in many
interacting-dark-energy growth analyses \cite{Komatsu:2014vna,Basilakos:2014tha,Sheykhi2022,Basilakos2024}, yielding
\begin{widetext}
\begin{equation}
\delta_i'' 
+ \left[ \frac{3}{a} + \frac{1}{H}\dv{H}{a} - \frac{1}{1+w_i}\dv{w_i}{a} \right]\delta_i'
+ \frac{w_i k^2}{a^4 H^2}\delta_i
= \frac{4\pi G (1+w_i)}{a^2 H^2}\sum_j (1+3w_j)\,\bar{\rho}_j \delta_j,
\label{eq:delta_special_case}
\end{equation}
\end{widetext}
here $i$ runs over the evolved components ($i=M,R$), while the entropic sector contributes only through the background $H(a)$.
\paragraph*{Fourier–space form and conventions used numerically.}
For numerical work we solve the Fourier–space version of Eq.~\eqref{eq:delta_special_case}, using
$\nabla^2\!\to\!-k^2$ and evolving the coupled system for $i=M$ (matter) and $i=R$ (radiation).
With $c_{s,i}^2=w_i(a)$ (adiabatic barotropic choice) the equations take the compact form
\begin{equation}
\delta_i''(k,a)\;=\;-A_i(a)\,\delta_i'(k,a)\;-\;C_i(k,a)\,\delta_i(k,a)\;+\;S_i(a),
\label{eq:code_compact}
\end{equation}
with
\begin{align}
A_i(a)&=\frac{3}{a}+\frac{H'}{H}-\frac{w_i'}{1+w_i},\\
C_i(k,a)&=\frac{w_i\,k^2}{a^4H^2},\\
S_i(a)&=\frac{4\pi G}{a^2H^2}\,(1+w_i)\sum_{j}\bigl(1+3w_j\bigr)\,\bar\rho_j\,\delta_j.
\label{eq:code_coeffs}
\end{align}
This is exactly the structure implemented in our Python integrator: the “friction” coefficient $A_i$ coincides with the one multiplying
$\delta_i'$ in Eq.~\eqref{eq:delta_special_case}, the gradient term $C_i$ reproduces the $(w_i k^2/a^4H^2)\,\delta_i$ contribution, and $S_i$ matches the Poisson source with the $(1+3w_j)$ weights. 
This is the standard subhorizon Newtonian formulation used in the growth–of–structure literature and in entropic/interactive backgrounds, reducing to the familiar friction structure when exchange terms are negligible \cite{Lima1997,Reis2003,Jesus2011,Komatsu:2014vna,Basilakos:2014tha}.
For cold matter $w_M=0$ the gradient term vanishes (so the late–time growth is $k$–independent), whereas for radiation $w_R=1/3$ the $k$–term is retained. Throughout we work on subhorizon scales ($k\gg aH$), neglect anisotropic stress, and use $\Phi=\Psi$.

\paragraph*{Background and effective equations of state.}
All background functions—$\bar\rho_i(a)$, $H(a)$ and their derivatives—are computed self–consistently from the MHEC background with parameters $(n,\gamma,\dots)$.
The effective equations of state $w_{M,\mathrm{eff}}(a)$ and $w_{R,\mathrm{eff}}(a)$ entering Eqs.~\eqref{eq:code_coeffs} include the interaction terms implied by the entropic sector; their derivatives $w_i'(a)$ are obtained by differentiating the same background solution. This ensures that the only modification to linear growth enters through the background expansion and the (time–dependent) effective $w_i(a)$, without introducing any ad hoc growth–sector parameters or functions (e.g.\ $\mu(a,k)$ or a tuned $\gamma_{\rm growth}$). This is analogous in spirit to dissipative/creation entropic growth analyses and to nonextensive (Tsallis/Barrow) entropic frameworks where friction and an effective clustering strength are modified by the background \cite{Komatsu:2014vna,Sheykhi2022,Basilakos2024}.

\paragraph*{Initial conditions and the $f\sigma_8$ observable.}
We initialize deep in the radiation era at $a_{\rm ini}\simeq 10^{-5}$ with adiabatic conditions
\[
\delta_R(a_{\rm ini})=\tfrac{4}{3}\,\delta_M(a_{\rm ini}),\qquad
\delta_R'(a_{\rm ini})=\tfrac{4}{3}\,\delta_M'(a_{\rm ini}),
\]
and evolve to $a=1$ using an adaptive ODE solver.
The growth rate and clustering amplitude are then obtained as
\[
f(a)=a\,\frac{\delta_M'(a)}{\delta_M(a)},\qquad
\sigma_8(a)=\sigma_{8,0}\,\frac{\delta_M(a)}{\delta_M(1)},
\]
and the observable used in the RSD comparison is $f\sigma_8(a)=f(a)\,\sigma_8(a)$ \cite{Sagredo2018}.
Because $w_M=0$ at late times, $f\sigma_8$ is scale–independent in our setup; a representative linear mode $k$ is used only to account for the radiation gradient term and does not affect the matter–sector prediction at the redshifts of interest.

\paragraph*{Remark on variable choices.}
Eq.~\eqref{eq:delta_special_case} is written in derivatives with respect to $a$. In terms of $\ln a$ one recovers the familiar “$2+\mathrm{d}\ln H/\mathrm{d}\ln a$” friction structure (modulo the small correction $-\mathrm{d}\ln(1+w_i)/\mathrm{d}\ln a$ that is explicit in $A_i$), making transparent the connection with standard $\Lambda$CDM growth when $w_M=0$ and $H(a)$ reduces to its concordance form.  In entropic/interactive formulations one often finds an additional background–exchange term $Q$ entering the friction and source pieces; in the limit where $Q/H\ll 1$ the equations reduce to the form used here \cite{Basilakos:2014tha}.

As shown in Appendix~\ref{appx}, the interaction-induced friction term $Q_m/\rho_m$
appears explicitly in the fully covariant formulation, implying a correction to the effective
friction in the growth equation controlled by $q_m(z)\equiv Q_m/(H\rho_m)$.

The numerical evaluation presented in Appendix~\ref{appx} (Fig.~\ref{fig:qm_full}),
based on the self-consistent MHEC background evolution (without invoking the $\Lambda$CDM approximation),
demonstrates that the interaction contribution is strongly suppressed in the parameter range relevant for our analysis.
In particular, already for $\log_{10}\gamma=-4$ one finds $|q_m(z)|<10^{-3}$ throughout $z\in[0,2]$,
while for more negative values of $\gamma$ the correction decreases rapidly.

Therefore, Eq.~(3.4) used in the main text corresponds to the leading-order sub-horizon limit
of the exact relativistic result, with the interaction-dependent correction remaining numerically subleading
over the considered redshift range.

\section{Data and Methodology}
The models are evaluated against both geometrical and dynamical probes. We retain the structure outlined below and detail the exact datasets used in our MCMC implementation.

\paragraph*{Type Ia supernovae (SNe~Ia).}
We employ distance moduli derived from 1701 light curves corresponding to 1550 spectroscopically confirmed Type Ia supernovae (SNe Ia) from the Pantheon+ compilation\footnote{https://github.com/PantheonPlusSH0ES/DataRelease} \cite{Brout:2022vxf}, spanning the redshift interval $0.001<z<2.26$. The supernova contribution to the $\chi^2$ statistic is defined as
%\[
$\chi^2_{\rm SN} \;=\; \Delta \boldsymbol{\mu}^{\rm SN} \cdot \mathbf{C}^{-1}_{\rm SN} \cdot \Delta \boldsymbol{\mu}^{\rm SN}$,
%\]
where $\Delta\boldsymbol{\mu} = \boldsymbol{\mu}_{\rm theo} - \boldsymbol{\mu}_{\rm obs}$ denotes the vector of residuals between the theoretical and observed distance moduli for each SN Ia, and $\mathbf{C}_{\rm SN}$ is the total covariance matrix, incorporating both statistical and systematic uncertainties.
The theoretical distance modulus is given by
\[
\mu_{\rm theo}(z_{\rm hel}, z_{\rm HD}, \boldsymbol{p}) = 25 + 5 \log_{10}\!\big[ d_{L}(z_{\rm hel}, z_{\rm HD}, \boldsymbol{p}) \big],
\]
where $d_L$ is the luminosity distance in Mpc, defined as
\begin{equation}
d_L(z_{\rm hel}, z_{\rm HD}, \boldsymbol{p}) \;=\; (1+z_{\rm hel}) \int_{0}^{z_{\rm HD}} \frac{c\,dz'}{H(z', \boldsymbol{p})},
\end{equation}
with $z_{\rm hel}$ the heliocentric redshift, $z_{\rm HD}$ the Hubble diagram redshift \cite{Carr:2021lcj}, and $\boldsymbol{p}$ the vector of cosmological parameters.
The observed distance modulus is
%\[
$\mu_{\rm obs} = m_{B} - M$,
%\]
where $m_{B}$ is the standardized rest-frame $B$-band apparent magnitude of the SN Ia and $M$ is the fiducial absolute magnitude, calibrated using primary distance indicators such as Cepheid variables. In analyses based solely on SNe Ia, $H_0$ and $M$ are normally degenerate. However, the Pantheon+ sample contains 77 SNe Ia hosted in galaxies with Cepheid-based distance measurements, which serve as external distance anchors and thereby break this degeneracy, enabling independent constraints on $H_0$ and $M$.
Consequently, the residual vector $\Delta\boldsymbol{\mu}$ takes the form
\begin{equation}
\Delta\boldsymbol{\mu}_i = 
\begin{cases}
    m_{B,i} - M - \mu_{{\rm Ceph},i}, & i \in \text{Cepheid hosts},\\[4pt]
    m_{B,i} - M - \mu_{{\rm theo},i}, & \text{otherwise},
\end{cases}
\end{equation}
where $\mu_{\rm Ceph}$ denotes the Cepheid-calibrated host-galaxy distance modulus provided by the Pantheon+ collaboration.

\paragraph*{CMB compressed likelihood.}
We use a four-parameter compressed CMB vector
\begin{equation}
\mathbf{v}_{\rm CMB} \equiv \bigl(R,\,\ell_a,\,\Omega_b h^2,\,[\Omega_m-\Omega_b]h^2\bigr),
\end{equation}
with the \emph{Planck} 2018-inspired mean and covariance (as in \cite{Wang:2007mza,Zhai:2019nad}), and with the model predictions computed from our background:
\begin{align}
R(\boldsymbol{p}) &\equiv \sqrt{\Omega_m H_0^2}\,\frac{r(z_\ast,\boldsymbol{p})}{c}, \\
\ell_a(\boldsymbol{p}) &\equiv \pi\,\frac{r(z_\ast,\boldsymbol{p})}{r_s(z_\ast,\boldsymbol{p})}, \\
r(z,\boldsymbol{p}) &= \int_{0}^{z} \frac{c\,dz'}{H(z',\boldsymbol{p})}, \\
r_s(z,\boldsymbol{p})&=\int_{z}^{\infty} \frac{c_s(z')\,dz'}{H(z',\boldsymbol{p})}.
\end{align}
We adopt updated fitting formulas for $z_\ast$ and the drag redshift $z_d$ \cite{Aizpuru2021,Hu:1995en}; the sound speed is
\begin{align} \label{eq:sound_speed}
c_s(z)=\frac{c}{\sqrt{3\bigl(1+\overline{R}_b(1+z)^{-1}\bigr)}},\\
\overline{R}_b=31500\,\Omega_b h^2\,(T_{\rm CMB}/2.7)^{-4},
\end{align}
with $T_{\rm CMB}=2.726~\mathrm{K}$. In our generalized entropic background, $H(z)$, $r(z)$ and the integrals for $r_s$ are computed self-consistently from Eqs.~(\ref{eq:Fa11})–(\ref{eq:weff_2}).
The expression for the sound speed, Eq.~(\ref{eq:sound_speed}), must be generalized, as it is only valid under the assumption that baryons scale as $\propto a^{-3}$ and radiation as $\propto a^{-4}$, i.e. when their equations-of-state parameters are fixed to the standard values $w_i = 0$ and $w_i = 1/3$, respectively. In our entropic cosmologies, however, the continuity equations are modified through \textit{effective} equations-of-state parameters, $w_{\mathrm{eff},i}$, which may deviate from these canonical values. The baryon-to-photon ratio is defined as
\begin{equation}
R_b \equiv \frac{\rho_b + p_b}{\rho_\gamma + p_\gamma} = \frac{\rho_b(1+w_b)}{\rho_\gamma(1+w_\gamma)}\,,
\end{equation}
which, for $w_b = 0$ and $w_\gamma = 1/3$, reduces to the standard form
\begin{equation}
R_b = \frac{3}{4}\frac{\rho_{b,0} (1+z)^3}{\rho_{\gamma,0}(1+z)^4} = \frac{3}{4}\frac{\Omega_b}{\Omega_\gamma} (1+z)^{-1},
\end{equation}
with $\overline{R}_{b}$ and its numerical prefactors determined from
$\Omega_\gamma = \Omega_{r}/(1+0.2271\,N_{\mathrm{eff}}) \approx 2.469 \times 10^{-5} h^{-2}$ \cite{WMAP:2008lyn}.
In the context of our entropic models, the baryon-to-photon ratio must instead be expressed in the more general form
\begin{equation}
R_b = \frac{(1+w_{\mathrm{eff},b})}{(1+w_{\mathrm{eff},\gamma})}\frac{\Omega_b}{\Omega_\gamma} \frac{\mathcal{F}_{b}(z)}{\mathcal{F}_{\gamma}(z)}\,,
\end{equation}
where $w_{\mathrm{eff},b}$ and $w_{\mathrm{eff},\gamma}$ are given by Eq. (\ref{eq:weff_2}). In this framework, the redshift dependence of the energy densities cannot be expressed in closed analytical form; instead, it must be obtained by numerically solving the coupled systems of continuity equations Eqs.~(\ref{eq:continuity_Lambda_MLn_2_entropy})–(\ref{eq:continuity_Lambda_MLn_mat_rad_2}). 
\smallskip
\paragraph*{BAO: DESI-DR2.}
In addition to the legacy BAO/RSD blocks described below, our baseline chains include the baryon acoustic oscillation measurements from DESI Data Release~2 \cite{AbdulKarim2025}. We use the published $D_M(z)/r_d$ and $D_H(z)/r_d$ pairs for the LRG, ELG and QSO samples at $z=\{0.510,\,0.706,\,0.934,\,1.321,\,1.484,\,2.330\}$, together with the $D_V(z)/r_d$ point at $z=0.295$ from the bright-galaxy sample, and their associated $2\times 2$ (or $1\times 1$) covariance matrices. The model predictions are obtained from our entropic background via
\begin{align}
D_M(z) &= \int_0^z \frac{c\,dz'}{H(z')},\\
D_H(z) &= \frac{c}{H(z)},\\
D_V(z) &\equiv \Bigl[(1+z)^2 D_A^2(z)\,\frac{c\,z}{H(z)}\Bigr]^{1/3},
\end{align}
and rescaled by the sound horizon at the drag epoch $r_d \equiv r_s(z_d)$ computed self-consistently using the same $z_d$ fit as in
the CMB analysis. These contributions enter the likelihood as $\chi^2_{\rm DESI}=\chi^2_{\rm LRG1}+\chi^2_{\rm LRG2}+\chi^2_{\rm LRG3/ELG1}+\chi^2_{\rm ELG2}+\chi^2_{\rm QSO(DESI)}+\chi^2_{\rm Ly\alpha}+\chi^2_{\rm BGS}$.
\smallskip
\paragraph*{Growth–rate data $f\sigma_{8}(z)$ and likelihood.}
We confront the linear–growth sector of the model with (i) the internally–validated compilation of $f\sigma_{8}(z)$ measurements assembled in \cite{Sagredo2018}, and (ii) the earlier \cite{Nesseris2017}, which introduced the curated \emph{Gold-2017} subset and the standard correction for survey–fiducial cosmologies. In this work we adopt the PRD-2018 set—namely Gold-2017 plus the SDSS-IV updates and their published covariance blocks—because it (a) performs a Bayesian “internal robustness” analysis that finds no anomalous subsets in current $f\sigma_{8}$ data, and (b) provides the explicit sub–covariances for the WiggleZ triplet and SDSS-IV quartet used in our likelihood. \cite{Sagredo2018,Nesseris2017}

\smallskip
\noindent\textit{Observable and definitions.} We use the standard, bias–independent combination
\begin{align}
f\sigma_{8}(a) &\equiv f(a)\,\sigma_{8}(a), \label{eq:fs8_def1}\\
f(a) &\equiv \frac{d\ln\delta_{m}}{d\ln a}, \label{eq:fs8_def2}\\
\sigma_{8}(a) &= \sigma_{8,0}\,D(a), \label{eq:fs8_def3}\\
D(a) &\equiv \frac{\delta_{m}(a)}{\delta_{m}(1)}. \label{eq:fs8_def4}
\end{align}
Equivalently,
\begin{equation}
f\sigma_{8}(a)=a\,\sigma_{8,0}\,D'(a),
\label{eq:fs8_alt}
\end{equation}
where a prime denotes $d/da$ and $\sigma_{8,0}$ is sampled as a free amplitude parameter.

\smallskip
\noindent\textit{Growth equation with the MHEC background.}
The background $H(a)$ and densities $\rho_{i}(a)$ follow  Eqs.~(\ref{eq:Fa11})–(\ref{eq:weff_2}). On sub–horizon scales we evolve the coupled matter–radiation perturbations using the Fourier–space system derived in Sec.~\ref{sec:lin_pert}, namely Eqs.~(\ref{eq:code_compact})–(\ref{eq:code_coeffs}) for $i=M,R$, evaluated at a fixed comoving wavenumber $k=0.002\,h\,{\rm Mpc}^{-1}$.
From the resulting matter contrast $\delta_{M}(a)$ we construct 
\begin{equation}
D(a)\equiv\frac{\delta_{M}(a)}{\delta_{M}(1)},\qquad
f(a)=\frac{d\ln\delta_{M}}{d\ln a}
=a\,\frac{\delta_{M}'(a)}{\delta_{M}(a)},
\end{equation}
so that $f\sigma_{8}(a)=f(a)\,\sigma_{8,0}D(a)$ as in Eqs.~(\ref{eq:fs8_def1})–(\ref{eq:fs8_alt}). In the limit $w_{M,\mathrm{eff}}\to 0$ and negligible radiation, this system reduces to the usual GR single–fluid growth equation.

\smallskip
\noindent\textit{Fiducial–cosmology rescaling.} Because published $f\sigma_{8}$ values assume survey–specific fiducial backgrounds, we rescale the theory at each $z_{i}$ using
\begin{equation}
f\sigma_{8}^{\mathrm{mr}}(z_{i})
= f\sigma_{8}^{\mathrm{th}}(z_{i})\,
\frac{H_{\mathrm{fid}}(z_{i})\,D_{A,\mathrm{fid}}(z_{i})}
{H_{\mathrm{model}}(z_{i})\,D_{A,\mathrm{model}}(z_{i})},
\label{eq:fs8_rescale_fix}
\end{equation}
with the (separately defined) angular–diameter distance
\begin{equation}
D_{A}(z)=\frac{1}{1+z}\int_{0}^{z}\frac{c\,dz'}{H(z')}\,.
\label{eq:DA_fix}
\end{equation}

\paragraph*{BAO\,+\,RSD block I: WiggleZ (three redshifts).}
We include the WiggleZ measurements at $z=\{0.44,0.60,0.73\}$~\cite{Blake:2012pj}, using the 9-component data vector
\begin{equation}
\begin{aligned}
\mathbf{d}_{\rm WiggleZ}=\bigl(&A(z_1),A(z_2),A(z_3),F(z_1),F(z_2),F(z_3),\\
& f\sigma_8(z_1),f\sigma_8(z_2),f\sigma_8(z_3)\bigr)^{\!\top},
\end{aligned}
\end{equation}
with the published $9\times 9$ covariance. We model the BAO observables as
\begin{align}
A(z) &= \frac{100\,h\,\sqrt{\Omega_m}\,D_V(z)}{c\,z},\\
D_V(z) &\equiv \Bigl[(1+z)^2 D_A^2(z)\,\frac{c\,z}{H(z)}\Bigr]^{1/3},\\
F(z) &= \frac{(1+z)\,D_A(z)\,H(z)}{c},\\
D_A(z)&=\frac{1}{1+z}\int_{0}^{z}\frac{c\,dz'}{H(z')},
\end{align}
while $f\sigma_8(z)$ is computed from linear growth as described above and mapped to the survey fiducial via the standard factor $[H_{\rm fid}D_{A,\rm fid}]/[H D_A]$.

\paragraph*{BAO\,+\,RSD block II: SDSS-IV DR14 QSO (four redshifts).}
We include the SDSS-IV DR14 quasar measurements at $z=\{0.978,\,1.230,\,1.526,\,1.944\}$~\cite{eBOSS:2018yfg}. The $12$-component vector stacks
\begin{equation}
\begin{aligned}
\mathbf{d}_{\rm QSO}=\bigl(&D_A(z_1),H(z_1),f\sigma_8(z_1),\ldots,\\
& D_A(z_4),H(z_4),f\sigma_8(z_4)\bigr)^{\!\top},
\end{aligned}
\end{equation}
with the full published $12\times 12$ covariance. Following~\cite{eBOSS:2018yfg}, we work with the \emph{rescaled combinations}
\begin{equation}
\begin{aligned}
&\frac{D_A(z)}{r_s(z_d)}\,r_{\rm fid}, \qquad 
H(z)\,\frac{r_s(z_d)}{r_{\rm fid}},\\
&\text{with }~r_{\rm fid}=147.78~\mathrm{Mpc},
\end{aligned}
\end{equation}
and $r_s(z_d)$ the model sound horizon at the baryon-drag epoch, computed self-consistently (same $z_d$ fit as above). This preserves the correlation between geometry and growth carried by the QSO sample. As for WiggleZ, the RSD observable $f\sigma_8(z)$ is mapped to the survey fiducial via the standard factor $[H_{\rm fid}D_{A,\rm fid}]/[H D_A]$ used in the likelihood.

\smallskip
\paragraph*{Uncorrelated growth set ($f\sigma_8$ only).}
Beyond the two joint blocks, we use the internally robust, uncorrelated RSD points compiled in \cite{Sagredo:2018ahx}, with cross-checks against the earlier \cite{Nesseris:2017vor}. These measurements span $z\simeq 0.02$–$1.4$ and we adopt the survey-specific fiducial $\Omega_{m,{\rm fid}}$ values to apply the Alcock–Paczyński rescaling.
% \begin{equation}
% \bigl[f\sigma_8\bigr]_{\rm map}(z_i)=\bigl[f\sigma_8\bigr]_{\rm th}(z_i)\,
% \frac{H_{\rm fid}(z_i)\,D_{A,\rm fid}(z_i)}{H(z_i)\,D_A(z_i)}.
% \end{equation}
The resulting $\chi^2_{f\sigma_8}$ is a diagonal sum over these points. See \cite{Sagredo2018,Nesseris2017} for the curation and validation of this set.
\smallskip
\paragraph*{Likelihood combination.}
Denoting by $\chi^2_{\rm SN}$ the Pantheon+ \& SH0ES contribution, by $\chi^2_{\rm CMB}$ the compressed CMB term, by $\chi^2_{\rm DESI}$ the sum of the DESI-DR2 BAO pieces (LRG, ELG, QSO, Ly$\alpha$, BGS), by $\chi^2_{\rm Wig}$ and $\chi^2_{\rm QSO}$ the quadratic forms built with the published covariance matrices of the WiggleZ and DR14-QSO BAO+RSD blocks, and by $\chi^2_{f\sigma_8}$ the diagonal $\chi^2$ of the remaining, uncorrelated growth–rate points, our total likelihood reads
\begin{equation}
-2\ln\mathcal{L}_{\rm tot}
=\chi^2_{\rm SN}
+\chi^2_{\rm CMB}
+\chi^2_{\rm DESI}
+\chi^2_{\rm Wig}
+\chi^2_{\rm QSO}
+\chi^2_{f\sigma_8}\,.
\end{equation}
In our MCMC analysis we do not vary $n$ and $\gamma$ simultaneously. Instead, we perform two families of runs. In the first family we fix $n$ to a set of representative values and sample
\begin{equation}
\{\Omega_m,\Omega_b,h,\log_{10}\gamma,M,\sigma_{8,0}\}
\end{equation}
under broad, uniform priors. In the second family we fix $\log_{10}\gamma$ to selected values and sample
\begin{equation}
\{\Omega_m,\Omega_b,h,n,M,\sigma_{8,0}\}\,.
\end{equation}
The total $\chi^2$ is minimized using \texttt{emcee} \cite{emcee}, a pure-Python implementation of the affine invariant MCMC method. We assess convergence and efficiency by following \cite{emcee, zbMATH05709093}, using integrated autocorrelation time ($\tau$) to estimate independent samples and trace plots to inspect walker stability. The acceptance fraction is maintained between $0.2\text{--}0.5$. We use 50 walkers with a burn-in of 1000 steps and 5000 steps for analysis. A mix of \texttt{StretchMove}, \texttt{DEMove}, and \texttt{KDEMove} proposal moves is used. For details, see \cite{emcee} and its documentation.

We calculate Bayesian evidence with \textsc{MCEvidence} \cite{Heavens:2017afp} to compare models $\mathcal{M}_1$ and $\mathcal{M}_2$. The Bayes factor, $B_{12} = \frac{\mathcal{Z}_2}{\mathcal{Z}_1}$, is interpreted using $\ln B_{12}=\Delta \ln Z$: negative values favor $\mathcal{M}_1$, positive favor $\mathcal{M}_2$, based on Jeffreys scale \cite{Jeffreys:1961, Kass_Raftery_1995, Trotta:2008qt}.

\section{Results}\label{sec:results}
In order to quantify how strongly current data constrain the generalized mass–to–horizon entropic cosmology, we performed two complementary sets of Bayesian model–comparison runs. In the first family, the entropy index $n$ was fixed to a discrete set of values while $\log_{10}\gamma$ and the standard cosmological parameters were sampled. In the second family, we instead fixed $\log_{10}\gamma$ and let $n$ vary. The corresponding posterior means and credible intervals are reported in Tables~\ref{tab:params} and \ref{tab:params2}, while Tables~\ref{tab:evidence_fixed_n} and \ref{tab:evidence_fixed_gamma} summarize the Bayesian evidences relative to $\Lambda$CDM.

For the runs with fixed $n$, Table~\ref{tab:params} shows that the standard background parameters are remarkably stable across the whole range $n = 0.5,1,1.5,2,2.5$. The matter density parameter $\Omega_m = 0.2928 \pm 0.0035$ remains remarkably stable across all models, showing no dependence on the choice of $n$. Similarly, the baryon density $\Omega_b$ exhibits minimal variation, ranging from $0.04724 \pm 0.00041$ (for $n=2.5$) to $0.04726 \pm 0.00042$ (for $n=0.5$), consistent with the $\Lambda$CDM value of $0.04716 \pm 0.00041$. The Hubble parameter $h$ shows slight tension between the mass-to-horizon models ($h \approx 0.6884 \pm 0.003$) and $\Lambda$CDM ($h = 0.6891 \pm 0.0029$), though the difference is within $1\sigma$. The coupling parameter $\log_{10} \gamma$ is poorly constrained across all power-law models, with values ranging from $-22 \pm 10$ to $-24.3 \pm 9.1$, indicating weak sensitivity of current data to this parameter. The absolute magnitude of Type Ia supernovae $M$ and the amplitude of matter fluctuations $\sigma_8 = 0.788$--$0.789 \pm 0.025$ are essentially indistinguishable between all models. Despite the near-degeneracy in parameter constraints, Bayesian model comparison (Table~\ref{tab:evidence_fixed_n}) using MCEvidence reveals a consistent preference for the mass-to-horizon entropic cosmologies over $\Lambda$CDM, with $\Delta \ln Z$ ranging from $+2.85$ to $+2.98$ across all values of $n$. According to the Jeffreys scale, these positive evidence values correspond to ``slight'' to ``moderate'' favor for the mass-to-horizon entropic models, suggesting that the additional parameter $\gamma$ provides a marginally better fit to the data despite the penalty imposed by increased model complexity. Notably, the preference shows weak dependence on $n$, with $n=1$ yielding the highest evidence ($\Delta \ln Z = 2.98$) and $n=2.5$ the lowest ($\Delta \ln Z = 2.85$), though all models remain statistically comparable within their uncertainties.

To investigate the role of the coupling strength more directly, we perform an alternative analysis by fixing $\log_{10} \gamma$ to discrete values ($-2$, $-4$, $-8$, $-12$, and $-16$) while allowing the mass-to-horizon index $n$ to vary freely. This approach reveals significant sensitivity of cosmological parameters to the coupling strength (Table~\ref{tab:params2}). For strong coupling ($\log_{10} \gamma = -2$), the model exhibits substantial deviations from $\Lambda$CDM: $\Omega_m = 0.2954^{+0.0043}_{-0.0037}$, $\Omega_b = 0.05652 \pm 0.00059$, and notably $h = 0.6209 \pm 0.0032$, which is approximately $7\sigma$ lower than the $\Lambda$CDM value. The mass-to-horizon index is tightly constrained to $n = 0.794^{+0.013}_{-0.015}$ in this regime. As the coupling weakens, parameter values converge toward $\Lambda$CDM, with $\Omega_b$ decreasing from $0.05652 \pm 0.00059$ ($\log_{10} \gamma = -2$) to $\sim 0.0466$--$0.0467$ ($\log_{10} \gamma \leq -8$), and $h$ increasing from $0.6209$ to $\sim 0.693$--$0.694$. Simultaneously, the constraint on $n$ degrades substantially, broadening from $n = 0.794^{+0.013}_{-0.015}$ ($\log \gamma = -2$) to $n = 2.2 \pm 1.1$ ($\log \gamma = -16$), reflecting the diminishing impact of the generalized mass-to-horizon relation. The Bayesian evidence analysis (Table~\ref{tab:evidence_fixed_gamma}) demonstrates a clear preference hierarchy: strong coupling is strongly disfavored ($\log_{10} \gamma = -2$: $\Delta \ln Z = -99.37$), moderate coupling is slightly disfavored ($\log_{10} \gamma = -4$: $\Delta \ln Z = -1.58$), while weak coupling regimes are moderately favored over $\Lambda$CDM ($\log_{10} \gamma = -8$, $-12$, $-16$: $\Delta \ln Z = +3.13$ to $+3.77$). This pattern indicates that current data prefer small values of $\gamma$, with the optimal coupling strength lying in the range $\log \gamma \lesssim -8$, where modifications are sufficiently subtle to accommodate observations while providing improved statistical fits compared to the standard cosmological model.

In summary, our comprehensive analysis of mass-to-horizon entropic cosmological models through both fixed-$n$ and fixed-$\gamma$ approaches reveals complementary insights into the viable parameter space and observational constraints. When fixing the mass-to-horizon index $n$, we find that all examined values ($n = 0.5$--$2.5$) yield nearly identical cosmological parameters and are uniformly favored over $\Lambda$CDM by $\Delta \ln Z \approx +2.9$, indicating a slight to moderate statistical preference that is largely insensitive to the functional form of the modification. However, the coupling parameter $\log_{10} \gamma$ remains poorly constrained in this regime ($\log_{10} \gamma \sim -22$ to $-24$ with uncertainties of $\pm 9$--$10$), suggesting that the data cannot effectively discriminate between different coupling strengths when $n$ is held fixed. Conversely, fixing $\gamma$ while allowing $n$ to vary reveals a strong dependence of both parameter constraints and model viability on the coupling strength. Strong coupling ($\log_{10} \gamma \gtrsim -4$) produces significant tensions with observations, particularly in $h$ and $\Omega_b$, and is statistically disfavored. Weak coupling regimes ($\log_{10} \gamma \lesssim -8$) emerge as the preferred scenario, offering moderate improvements over $\Lambda$CDM ($\Delta \ln Z \approx +3.1$ to $+3.8$) while maintaining parameter values consistent with standard cosmology. The optimal region appears to lie at $\log_{10} \gamma \lesssim -8$ with relatively unconstrained $n \gtrsim 1$, where modifications to general relativity are sufficiently subtle to evade current observational bounds yet provide statistically meaningful improvements to the cosmological fit. These results suggest that future high-precision observations, particularly those targeting the Hubble tension and baryon abundance measurements, will be crucial for definitively testing weak-field modifications of gravity and potentially breaking the degeneracy between $n$ and $\gamma$ that currently limits our ability to distinguish between different theoretical implementations of modified gravity.

\begin{table*}
\caption{\label{tab:params}
Mean values and 68\% confidence limits for cosmological parameters
for different fixed values of $n$ and $\Lambda$CDM.}
\begin{ruledtabular}
\begin{tabular}{lcccccc}
Parameter & $n=0.5$ & $n=1$ & $n=1.5$ & $n=2$ & $n=2.5$ & $\Lambda$CDM \\
\hline

{\boldmath$\Omega_m       $} & $0.2928\pm 0.0035          $ & $0.2928\pm 0.0035          $ & $0.2928\pm 0.0035          $ & $0.2928\pm 0.0035          $ & $0.2928\pm 0.0035                   $ & $0.2928\pm 0.0035          $\\

{\boldmath$\Omega_b       $} & $0.04726\pm 0.00042        $ & $0.04725\pm 0.00041        $ & $0.04725\pm 0.00041        $ & $0.04725\pm 0.00041        $ & $0.04724\pm 0.00041               $ & $0.04716\pm 0.00041        $\\

{\boldmath$h              $} & $0.6882^{+0.0032}_{-0.0028}$ & $0.6884\pm 0.0029          $ & $0.6884\pm 0.0029          $ & $0.6884\pm 0.0029          $ & $0.6884\pm 0.0030                   $ & $0.6891\pm 0.0029          $\\

{\boldmath$\log_{10} \gamma    $} & $-22\pm 10                 $ & $-22\pm 10                 $ & $-23\pm 10                 $ & $-23.3\pm 9.5              $ & $-24.3\pm 9.1                       $ &      --                       \\

{\boldmath$M              $} & $-19.3997\pm 0.0095        $ & $-19.3991\pm 0.0088        $ & $-19.3993\pm 0.0089        $ & $-19.3992\pm 0.0088        $ & $-19.3992\pm 0.0089              $ & $-19.3971\pm 0.0088        $\\

{\boldmath$\sigma_8       $} & $0.788\pm 0.025            $ & $0.789\pm 0.025            $ & $0.788\pm 0.025            $ & $0.788\pm 0.025            $ & $0.788\pm 0.025                        $ & $0.789\pm 0.025            $\\
\end{tabular}
\end{ruledtabular}
\end{table*}

\begin{table*}
\caption{\label{tab:params2}
Mean values and 68\% confidence limits for cosmological parameters
for different fixed values of $\gamma$ and $\Lambda$CDM.}
\begin{ruledtabular}
\begin{tabular}{lcccccc}
Parameter & $\log_{10} \gamma=-2$ & $\log_{10} \gamma=-4$ & $\log_{10} \gamma=-8$  &$\log_{10} \gamma=-12$ & $\log_{10} \gamma=-16$ & $\Lambda$CDM \\
\hline
{\boldmath$\Omega_m       $} & $0.2954^{+0.0043}_{-0.0037}$ & $0.2929\pm 0.0036          $ & $0.2925\pm 0.0035          $ & $0.2921\pm 0.0036          $ & $0.2926\pm 0.0034          $ & $0.2928\pm 0.0035          $\\

{\boldmath$\Omega_b       $} & $0.05652\pm 0.00059        $ & $0.04738\pm 0.00042        $ & $0.04667\pm 0.00041        $ & $0.04658\pm 0.00044        $ & $0.04669^{+0.00037}_{-0.00043}$ & $0.04716\pm 0.00041        $\\

{\boldmath$h              $} & $0.6209\pm 0.0032          $ & $0.6873\pm 0.0031          $ & $0.6935\pm 0.0034          $ & $0.6944^{+0.0031}_{-0.0039}$ & $0.6932^{+0.0031}_{-0.0027}$ & $0.6891\pm 0.0029          $\\

{\boldmath$n              $} & $0.794^{+0.013}_{-0.015}   $ & $0.84^{+0.17}_{-0.24}      $ & $1.50^{+0.55}_{-0.90}      $ & $2.4^{+1.5}_{-1.6}         $ & $2.2\pm 1.1                $ & --                            \\

{\boldmath$M              $} & $-19.619\pm 0.011          $ & $-19.4027\pm 0.0092        $ & $-19.3834^{+0.0089}_{-0.010}$ & $-19.3808^{+0.0092}_{-0.012}$ & $-19.3843^{+0.0093}_{-0.0083}$ & $-19.3971\pm 0.0088        $\\

{\boldmath$\sigma_8       $} & $0.775\pm 0.024            $ & $0.788\pm 0.025            $ & $0.790^{+0.025}_{-0.028}   $ & $0.797^{+0.026}_{-0.031}   $ & $0.788\pm 0.025            $ & $0.789\pm 0.025            $\\
\end{tabular}
\end{ruledtabular}
\end{table*}

\begin{table}[h]
\centering
\caption{Bayesian evidence (log-evidence) from MCEvidence for fixed values of mass-to-horizon scaling index $n$. 
$\Delta \ln Z$ relative to $\Lambda$CDM.}
\label{tab:evidence}
\begin{tabular}{lcccc}
\hline\hline
Model & $\ln Z$ & $\sigma_{\ln Z}$ & $\Delta \ln Z$ & Interpretation \\
\hline
$\Lambda$CDM & $-820.8515$ & $0.0733$ & $0.0000$ & - \\
n = 0.5 & $-817.8946$ & $0.1163$ & $2.9569$ & Slightly favored \\
n = 1 & $-817.8710$ & $0.1157$ & $2.9806$ & Slightly favored \\
n = 1.5 & $-817.9011$ & $0.1128$ & $2.9505$ & Slightly favored \\
n = 2 & $-817.9497$ & $0.1091$ & $2.9018$ & Slightly favored \\
n = 2.5 & $-817.9991$ & $0.1165$ & $2.8524$ & Slightly favored \\
%n = 3 & $-817.8466$ & $0.1132$ & $3.0049$ & Moderately favored \\
\hline\hline
\end{tabular}
\label{tab:evidence_fixed_n}
\end{table}

\begin{table}[h]
\centering
\caption{Bayesian evidence (log-evidence) from MCEvidence for fixed $\gamma$ scenarios. 
$\Delta \ln Z$ relative to $\Lambda$CDM.}
\label{tab:evidence}
\begin{tabular}{lcccc}
\hline\hline
Model & $\ln Z$ & $\sigma_{\ln Z}$ & $\Delta \ln Z$ & Interpretation \\
\hline
$\Lambda$CDM & $-820.8515$ & $0.0733$ & $0.0000$ & -- \\
$\log_{10} \gamma$= -2 & $-920.2264$ & $0.1165$ & $-99.3748$ & Disfavored \\
$\log_{10} \gamma$= -4 & $-822.4269$ & $0.1113$ & $-1.5754$ & Slightly disfavored \\
$\log_{10} \gamma$= -8 & $-817.7232$ & $0.2184$ & $3.1283$ & Moderately favored \\
$\log_{10} \gamma$= -12 & $-817.0831$ & $0.2302$ & $3.7684$ & Moderately favored \\
$\log_{10} \gamma$= -16 & $-817.0831$ & $0.2302$ & $3.7684$ & Moderately favored \\
\hline\hline
\end{tabular}
\label{tab:evidence_fixed_gamma}

\end{table}

\section{Discussion}

\paragraph*{Comparison with earlier growth tests.}
Most entropic–cosmology constructions in the literature were confronted primarily with background probes; direct tests with the linear–growth observable $f\sigma_8(z)$ are comparatively rare. When performed, the \emph{canonical} entropic–force setups—typically driven by $H^2$ and/or $\dot H$ terms in the effective Friedmann equation but lacking an explicit constant term—were shown to be disfavored by structure–formation data unless they are driven toward a $\Lambda$CDM–like limit: the altered expansion history modifies the friction term $(2+\mathrm{d}\ln H/\mathrm{d}\ln a)$ in the growth equation and yields late–time clustering inconsistent with RSD unless the model effectively reintroduces a cosmological–constant–like contribution \cite{Basilakos:2014tha}. Dissipative/matter–creation variants (in which horizon thermodynamics sources irreversible entropy production) can accommodate the observed growth only for \emph{very small} creation/dissipation rates: increasing the rate enhances the effective friction and suppresses $\delta(a)$, producing too little growth at low redshift, whereas for $\tilde\mu\!\lesssim\!0.1$ the predicted $f\sigma_8$ can track the data \cite{Komatsu:2014vna}. Tsallis/Barrow–type nonextensive scenarios, by contrast, alter both the background and the effective clustering sector (often captured as a modified friction and a mild $G_{\rm eff}$ renormalization), and for suitable nonextensivity indices they can \emph{qualitatively} reproduce the measured $f\sigma_8(z)$ locus and soften the reported $\sigma_8$ tension; however, most of these works presented curve–overlays against binned RSD points rather than full covariance–level MCMC fits to the standard compilation \cite{Sheykhi2022,Basilakos2024}. 

Against this backdrop, our generalized mass–to–horizon entropic cosmology (MHEC) performs a \emph{direct} MCMC to the $f\sigma_8$ data used here (WiggleZ, BOSS, and related RSD points), jointly with the geometric set (SN, BAO, CMB), thus placing the growth sector on the same statistical footing as the background. The crucial ingredient is thermodynamic consistency: once the non–Bekenstein entropy is implemented together with the generalized mass–to–horizon prescription and the horizon temperature $T_H\!\propto\!1/L$, the modified $H(a)$ induces only mild, scale–independent changes in the linear–growth friction relative to $\Lambda$CDM, and no ad hoc extra parameters (e.g.\ a hand–tuned growth index or an explicit $\mu(a,k)$) are introduced. In this setup we obtain \emph{statistically competitive} fits across all redshift bins of the RSD compilation while remaining fully consistent with the geometric constraints, i.e.\ growth no longer stands in the way for entropic cosmology in a thermodynamically consistent framework \cite{Ali2025}. In addition, the Bayesian evidence analysis of Sec.~\ref{sec:results} shows that such MHEC realizations are, at most, weakly to moderately preferred over $\Lambda$CDM, with the data simultaneously constraining the entropic sector to remain close to the concordance limit.

In parallel to the entropic--force and Tsallis/Barrow nonextensive scenarios already discussed above, there is a growing body of work that studies structure formation in entropic or entropy–motivated dark energy models based on holographic principles. A first example is provided by Tsallis holographic dark energy (THDE), where the dark energy density is built from a nonadditive entropy--area relation for a cosmological horizon. In Ref.~\cite{daSilva:2021TsallisPert} da~Silva and Silva solved the full set of relativistic perturbation equations for several THDE realisations and constrained them with a combined data set of geometrical probes and $f\sigma_8$ measurements. They found that THDE models can fit the low--redshift growth data and mildly alleviate the $H_0$ and $\sigma_8$ tensions, but model selection criteria still tend to disfavour them with respect to $\Lambda$CDM. Conceptually, their strategy is close in spirit to ours: the entropic ingredients modify the background expansion, while the growth of matter perturbations is computed within standard general relativity. Our MHEC analysis extends this logic to a thermodynamically consistent entropic framework derived from a generalized mass--to--horizon relation, and confirms that such entropic deformations can remain compatible with current growth data while being only weakly to moderately preferred over $\Lambda$CDM according to the Bayes factors obtained here.

A complementary perspective is offered by Astashenok and Tepliakov, who analyzed the evolution of metric and matter perturbations in Tsallis holographic dark energy while explicitly treating the dark component as a boundary phenomenon rather than as an ordinary fluid~\cite{Astashenok:2024TsallisPert}. By perturbing the future event horizon (and, alternatively, a Hubble--scale cutoff) they showed that, for a wide range of Tsallis indices and cutoff parameters, both metric and dark energy perturbations either decay or freeze at late times, and remain under control even in the presence of matter--dark energy interaction. Their results indicate that holographic models based on nonadditive entropies need not suffer from catastrophic growth of dark energy inhomogeneities. This complements our working assumption of a smooth entropic component at sub--horizon scales: the MHEC background modifications can be embedded in a broader class of entropy--based models where the clustering sector remains perturbatively stable.

More generally, generalized nonextensive entropies have recently been implemented in a unified holographic dark energy (HDE) framework and tested against cosmological observations~\cite{Cimdiker:2025NonextensiveHDE}. In that analysis, Cimdiker, Dąbrowski and Salzano considered Barrow, Tsallis--Cirto, R\'enyi, Sharma--Mittal and Kaniadakis entropies as alternative holographic screens, and constrained the corresponding HDE models with background--level data. They found that all such nonextensive HDE variants are statistically disfavoured with respect to $\Lambda$CDM, and that the nearly extensive regime of the entropy parameters is observationally preferred. Our growth--of--structure constraints on the MHEC deformation parameter point in a similar qualitative direction: the data favour entropic modifications that remain close to the $\Lambda$CDM limit, reinforcing the picture in which nonextensive effects, if present, are relatively small at late times and in which strongly coupled entropic scenarios (such as $\log_{10}\gamma=-2$ in our analysis) are decisively excluded.

Finally, two recent works have applied the same generalized mass--to--horizon entropy that underlies our MHEC model to independent observational probes. Luciano and Paliathanasis confronted the generalized MHR cosmology with Type~Ia supernovae, cosmic chronometers and BAO data (including DESI~DR2), supplemented by the SH0ES prior on $H_0$~\cite{LucianoPaliathanasis:2025MHR_DESI}. They found that the entropic extension produces fits that are slightly better or statistically comparable to $\Lambda$CDM, with the $\Lambda$CDM limit lying well within the $1\sigma$ region of their constraints. In a follow--up work, Luciano studied the implications of the same framework for the growth of matter perturbations within the spherical top--hat formalism and for the primordial gravitational--wave background~\cite{Luciano:2026MHR_GrowthPGW}. There, the generalized MHR is shown to impact both the linear collapse history and the relic PGW spectrum, while still allowing parameter ranges consistent with current bounds. Our analysis complements these studies by performing a direct confrontation of the MHEC model with linear growth data in the standard $f\sigma_8$ language and by combining growth, SN, BAO and CMB--compressed information in a single global fit. Taken together, these results indicate that generalized mass--to--horizon entropic cosmologies form a coherent and thermodynamically well--motivated class of models in which the observed growth of cosmic structures can be accommodated without significant tension with $\Lambda$CDM, while Bayesian model comparison provides at most weak-to-moderate evidence in favour of \emph{small}, near--$\Lambda$CDM entropic corrections and no support for sizeable departures from the concordance paradigm.

The covariant analysis presented in Appendix \ref{appx} demonstrates that the interaction-induced perturbative terms are numerically subleading within the viable parameter space. The dominant deviation from $\Lambda$CDM arises from the modified background expansion $H(z)$, which is fully incorporated in the growth analysis.

Future high-precision Stage-IV surveys (Euclid, LSST, DESI) may reach the sensitivity required to probe these subleading covariant corrections.
\section{Conclusions}

We have presented a comprehensive observational analysis of generalized mass-to-horizon entropic cosmology (MHEC), extending previous background-only tests by incorporating the full suite of linear structure formation data alongside geometric probes. Our framework is built upon a thermodynamically consistent generalized mass-to-horizon relation $M \propto L^n$ and the corresponding entropy functional $S_n \propto L^{n+1}$, from which cosmic acceleration emerges as an entropic phenomenon driven by horizon thermodynamics. The key innovation of this approach lies in its restoration of thermodynamic consistency between generalized entropy functionals and the Hawking temperature, a consistency that was lacking in previous entropic cosmology constructions.

Our Bayesian analysis, combining Pantheon+ Type Ia supernovae with SH0ES calibration, DESI DR2 baryon acoustic oscillations, CMB distance priors, and redshift-space distortion measurements of $f\sigma_8(z)$, reveals several important findings. When fixing the mass-to-horizon scaling index $n$ and allowing the coupling parameter $\gamma$ to vary, all examined values yield nearly identical cosmological parameters that are statistically indistinguishable from $\Lambda$CDM, with standard background parameters remaining remarkably stable across all models. Bayesian model comparison consistently favors these MHEC realizations over $\Lambda$CDM, corresponding to slight-to-moderate preference on the Jeffreys scale, though the coupling parameter remains poorly constrained.

The complementary analysis with fixed $\gamma$ and varying $n$ provides crucial insights into the viable parameter space. Strong coupling regimes are decisively excluded, producing substantial tensions with observations and yielding Bayesian evidence strongly disfavoring such models. As the coupling weakens, the model predictions converge smoothly toward $\Lambda$CDM values, and for weak coupling the MHEC framework becomes moderately favored over the standard model. In this regime, the mass-to-horizon scaling index becomes increasingly unconstrained, reflecting the diminishing observational impact of the entropic sector.

A critical achievement of this work is the demonstration that thermodynamically consistent entropic cosmology can accommodate the observed growth of cosmic structures. Previous entropic-force constructions typically failed to reproduce the measured evolution of $f\sigma_8(z)$ without being driven back toward a $\Lambda$CDM-like limit. Our MHEC framework incorporates the growth data through direct MCMC fitting to the full covariance structure of joint BAO+RSD blocks and uncorrelated RSD compilations. The modified expansion history and effective equations of state induced by the entropic sector translate into scale-independent modifications of the linear growth friction term, without introducing ad hoc growth-sector parameters. The resulting predictions for $f\sigma_8(z)$ across all redshift bins remain fully consistent with observations, thereby resolving a long-standing tension in entropic cosmology.

The physical interpretation of our findings is clear. The weak-coupling regime that emerges as observationally preferred corresponds to scenarios where the entropic contribution behaves as a very slowly varying cosmological ``constant'' at late times. This provides a concrete realization of the entropic origin of dark energy: the observed cosmic acceleration can be attributed to horizon thermodynamics rather than to a fundamental cosmological constant, with the entropic sector effectively mimicking $\Lambda$ in the present epoch. The broad viability of the MHEC parameter space indicates that current observations cannot uniquely determine the functional form of the generalized entropy, pointing to the need for future high-precision measurements.

% Looking forward, upcoming Stage IV surveys will deliver percent-level measurements of the growth rate across wide redshift ranges, potentially breaking the current degeneracy between $n$ and $\gamma$ by probing the detailed redshift evolution of the entropic contribution. High-precision determinations of $H_0$ from independent anchors combined with improved CMB polarization data will tighten constraints on early-universe modifications induced by the entropic sector. Joint analyses incorporating weak lensing, cluster abundances, and velocity field reconstructions will test the consistency of MHEC predictions across multiple clustering observables, providing stringent checks on the assumption of a smooth entropic component at subhorizon scales.

In conclusion, generalized mass-to-horizon entropic cosmology emerges as a theoretically well-motivated and observationally viable framework for understanding the origin of cosmic acceleration. By restoring thermodynamic consistency through the generalized mass-to-horizon relation and confronting the model with the full range of geometric and dynamical probes, we have demonstrated that entropic forces on cosmological horizons can account for both the background expansion history and the growth of cosmic structures, while being statistically competitive with or mildly preferred over $\Lambda$CDM. Our analysis establishes that the observed late-time acceleration need not be attributed to a fundamental cosmological constant, but can instead arise naturally from the thermodynamics of horizons in an expanding universe. This entropic perspective opens new theoretical avenues for addressing foundational questions about the nature of dark energy and the deep connections between gravity, thermodynamics, and quantum information in cosmological contexts.
\appendix 

\section{Covariant Conservation and Linear Growth in the Entropic Framework} \label{appx}

In this framework the cosmological dynamics are governed by the modified Friedmann equations
\begin{align}
H^2 &= \frac{8\pi G}{3}\sum_{i=m,r} \rho_i
+  \frac{4 \pi G}{3} \rho_{n}  \left(3n-1\right) ,
\\
\frac{\ddot a}{a}
&=-\frac{4\pi G}{3}\sum_{i=m,r}\left(\rho_i+\frac{3p_i}{c^2}\right)
+  \frac{4 \pi G}{3} \rho_{n}  \left(3n-1\right) ,
\end{align}
where the index $i$ runs over matter and radiation only. 
The entropic density $\rho_n$ is not introduced as an independent thermodynamic fluid with a fundamental equation of state. Instead, it is a geometrically defined contribution arising from the entropy scaling, and is parametrized as
\begin{equation}
\rho_n(H)=\gamma\,\frac{c^{\,n-1}}{4\pi G}\,H^{3-n}.
\end{equation}
It follows that $\rho_n$ is fully determined by $H(t)$ and does not obey an independent continuity equation of the form
$\dot\rho_n+3H(1+w_n)\rho_n=0$.

General covariance requires $\nabla_\mu G^{\mu\nu}=0$, and therefore the total effective energy-momentum tensor must satisfy
\begin{equation}
\nabla_\mu T^{\mu\nu}_{\rm eff}=0.
\end{equation}
From the modified Friedmann equations one finds that the entropic sector contributes effectively as a vacuum-like term
\begin{equation}
T^{\mu\nu}_{n,{\rm eff}}
=
-\rho_{n,{\rm eff}}\, g^{\mu\nu},
\qquad
\rho_{n,{\rm eff}}=\frac{3n-1}{2}\rho_n,
\end{equation}
so that the conserved total energy density is
\begin{equation}
\rho_{\rm tot}
=
\sum_{i=m,r}\rho_i
+
\frac{3n-1}{2}\rho_n .
\end{equation}

Differentiating the Friedmann equation and using the acceleration equation to eliminate $\dot H$, one obtains the generalized conservation law
\begin{equation}
\sum_{i=m,r}\dot{\rho}_i
+
3H\sum_{i=m,r}\rho_i(1+w_i)
=
-\frac{3n-1}{2}\dot{\rho}_n .
\label{eq:global_conservation}
\end{equation}
Equation (\ref{eq:global_conservation}) shows that the standard components are not separately conserved; instead, the Bianchi identity enforces conservation of the combination $\sum_i\rho_i+\frac{3n-1}{2}\rho_n$.

To rewrite the system in interacting-fluid form, we introduce source functions $Q_m$ and $Q_r$ through
\begin{align}
\dot{\rho}_m+3H\rho_m &= Q_m,
\\
\dot{\rho}_r+4H\rho_r &= Q_r,
\end{align}
while the entropic density satisfies the constraint
\begin{equation}
\dot{\rho}_n
=
- A_m H^{2-n}\rho_m
- A_r H^{2-n}\rho_r .
\end{equation}
Consistency with Eq.~(\ref{eq:global_conservation}) requires
\begin{equation}
Q_i=\frac{3n-1}{2}A_i H^{2-n}\rho_i,
\qquad i=m,r.
\end{equation}
The coefficients $C_n^{\gamma}$, $A_m$ and $A_r$ given in Sec. \ref{sec:II}, are determined from the geometric definition of $\rho_n$.
% \begin{equation}
% C_n^{\gamma}=\frac{(3-n)\gamma c^{\,n-1}}{4\pi G},
% \qquad
% A_m=4\pi G C_n^{\gamma},
% \qquad
% A_r=\frac{16\pi G}{3}C_n^{\gamma}.
% \end{equation}
Thus the interaction terms are not postulated independently but arise from consistency between the entropic parametrization and the Bianchi identity.

At the perturbative level we work in Newtonian gauge,
\begin{equation}
ds^2=-(1+2\Psi)dt^2+a^2(1-2\Phi)d\vec{x}^2.
\end{equation}
For pressureless matter, the perturbed continuity equation in the presence of the interaction reads
\begin{equation}
\dot{\delta}_m
=
-\theta_m
+3\dot{\Phi}
+
\frac{\delta Q_m}{\rho_m}
-
\frac{Q_m}{\rho_m}\delta_m
-
\frac{Q_m}{\rho_m}\Psi ,
\end{equation}
where $\delta_m\equiv\delta\rho_m/\rho_m$ and $\theta_m$ is the velocity divergence. 
Perturbing $Q_m$ yields
\begin{equation}
\delta Q_m
=
Q_m \delta_m
+
\rho_m\,\delta\!\left(H^{2-n}\right)\frac{3n-1}{2}A_m .
\end{equation}
Since $\delta(H^{2-n})=(2-n)H^{1-n}\delta H$, the interaction induces additional metric contributions proportional to $\dot{\Phi}$.

The Euler equation becomes
\begin{equation}
\dot{\theta}_m
+
\left(H-\frac{Q_m}{\rho_m}\right)\theta_m
-
\frac{k^2}{a^2}\Psi
=0.
\end{equation}
Combining the perturbed continuity and Euler equations, and using the Poisson equation
\begin{equation}
\frac{k^2}{a^2}\Psi
=
4\pi G\rho_m\delta_m
\end{equation}
in the sub-horizon regime, one obtains the exact linear growth equation
\begin{equation}
\ddot{\delta}_m
+
\left(
2H
-
\frac{Q_m}{\rho_m}
\right)
\dot{\delta}_m
-
4\pi G\rho_m\delta_m
=
0,
\end{equation}
with
\begin{equation}
\frac{Q_m}{\rho_m}
=
\frac{3n-1}{2}A_m H^{2-n}.
\end{equation}
For $n=3$ one has $A_m=A_r=0$ and $\rho_n=\mathrm{const}$, so that the standard $\Lambda$CDM growth equation is exactly recovered. 

Therefore, in this entropic construction the matter and radiation sectors are not independently conserved; rather, energy exchange with the geometrically defined entropic component is enforced by the Bianchi identity. The conserved quantity is the effective combination $\sum_i\rho_i+\frac{3n-1}{2}\rho_n$, and the interaction terms $Q_m$ and $Q_r$ follow uniquely from the entropy scaling of $\rho_n(H)$.

\subsection{Self-consistent background evolution and numerical evaluation of $q_m(z)$}

In the previous subsections we derived the interaction term
\begin{equation}
\frac{Q_m}{\rho_m}
=
\frac{3n-1}{2}A_m H^{2-n},
\qquad
A_m=(3-n)\gamma,
\end{equation}
which follows uniquely from the modified Friedmann equations and the Bianchi identity.

In order to evaluate the magnitude of the interaction-induced friction term
in a fully self-consistent manner, we must compute the background expansion
without resorting to the $\Lambda$CDM approximation.

\paragraph{Modified Friedmann equation.}

From Eq.~(2.4) and the definition
\(
\rho_{n,\mathrm{eff}}=\frac{3n-1}{2}\rho_n
\),
the Friedmann equation reads
\begin{equation}
H^2
=
\frac{8\pi G}{3}\rho_m
+
\frac{4\pi G}{3}(3n-1)\rho_n(H).
\end{equation}

Using
\begin{equation}
\rho_n(H)=\gamma \frac{c^{\,n-1}}{4\pi G} H^{3-n},
\end{equation}
and introducing the dimensionless quantity
\begin{equation}
E(z)\equiv \frac{H(z)}{H_0},
\end{equation}
we obtain the algebraic equation
\begin{equation}
E^2(z)
=
\Omega_{m0}(1+z)^3
+
\frac{3n-1}{2}\gamma E^{3-n}(z).
\label{eq:background_algebraic}
\end{equation}

Equation (\ref{eq:background_algebraic}) is solved numerically
for each redshift $z$.
No $\Lambda$CDM approximation is employed.
The flatness condition is automatically satisfied through
the normalization $E(0)=1$.

\paragraph{Interaction parameter.}

Using the exact background solution,
the dimensionless interaction parameter is

\begin{equation}
q_m(z)
\equiv
\frac{Q_m}{H\rho_m}
=
\frac{3n-1}{2}(3-n)\gamma E^{1-n}(z).
\label{eq:qm_exact}
\end{equation}

Equation (\ref{eq:qm_exact}) is evaluated numerically
using the full solution of Eq.~(\ref{eq:background_algebraic}).

\paragraph{Numerical results.}

\begin{figure}[h]
\centering
\includegraphics[width=0.5\textwidth]{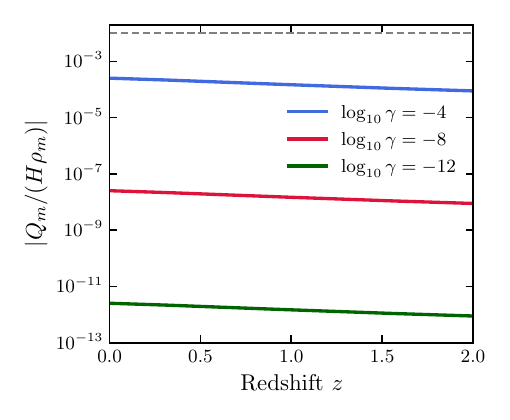}
\caption{
Magnitude of $q_m(z)=Q_m/(H\rho_m)$ computed using the full self-consistent MHEC background evolution
(Eq.~\ref{eq:background_algebraic}) for representative values $\log_{10}\gamma=-4,-8,-12$
with $\Omega_{m0}=0.2925$.
The horizontal dashed line marks the $1\%$ level.
}
\label{fig:qm_full}
\end{figure}

Figure~\ref{fig:qm_full} shows $|q_m(z)|$, where $q_m(z)\equiv Q_m/(H\rho_m)$,
computed using the full self-consistent MHEC background evolution
(Eq.~\ref{eq:background_algebraic}) for representative values
$\log_{10}\gamma=-4,-8,-12$ (with $n\simeq 2$ and $\Omega_{m0}=0.2925$).
The horizontal dashed line marks the $1\%$ level.
The case $\log_{10}\gamma=-2$ is omitted, since it leads to percent-level corrections at low redshift and is strongly disfavoured by the background constraints.

For $\log_{10}\gamma=-4$ the interaction-induced correction satisfies
\begin{equation}
|q_m(z)| < 10^{-3}
\quad \text{for } z\in[0,2],
\end{equation}
and it is further suppressed for smaller $\gamma$, consistent with the scaling $|q_m|\propto \gamma$
implied by Eq.~\ref{eq:qm_exact}.

\paragraph{Implication for the growth equation.}

The exact linear growth equation derived above,
\begin{equation}
\ddot{\delta}_m
+
\left(
2H
-
\frac{Q_m}{\rho_m}
\right)\dot{\delta}_m
-
4\pi G\rho_m\delta_m
=0,
\end{equation}
therefore contains a correction to the Hubble friction term
which remains below the percent level
throughout the viable parameter space favored by background data.

We conclude that,
although the model is formally non-conservative
and requires the covariant treatment presented in this Appendix,
the interaction-dependent perturbative term is numerically subleading.
The dominant physical effect of MHEC on structure formation
arises from the modified background expansion $H(z)$,
which is fully included in the analysis of the main text.

\bibliographystyle{apsrev4-1}
\bibliography{ref1}

\end{document}